\documentclass[twocolumn,showpacs,preprintnumbers,amsmath,amssymb]{revtex4}
\usepackage{graphicx}
\usepackage{epsfig}
\usepackage{dcolumn}
\usepackage{bm}
\def\bk{\hspace{-5.5pt}/}

\begin{document}

\title{Production of new chiral bosons at Tevatron and LHC}

\author{Mihail V. Chizhov}
\affiliation{Centre for Space Research and Technologies, Faculty of Physics,\\
University of Sofia, 1164 Sofia, Bulgaria
}%


\begin{abstract}
The possibility of production of new spin-1 chiral particles at
the hadron colliders the Tevatron and the LHC is considered. Their
coupling constants to ordinary fermions can be fixed using the
hypothesis about a dynamical generation of kinetic terms for the
bosons and assuming an universality of interactions of these
particles. The masses of the chiral particles can be predicted
from experimental data of precise low-energy experiments, which
already indicate indirectly the existence of such new particles.
Quantitative estimations of the production cross-sections of the
chiral particles are made. It is noteworthy that, due to their
non-gauge interactions with fermions, their decays do not lead to
the Jacobian peak in the transverse momentum/mass distribution.
Moreover, the lightest chiral particles show leptophobic
character, which  makes their detection at hadron colliders
challenging.
\end{abstract}


\maketitle

\section{Introduction}

The standard model (SM) contains and unifies interactions of three
different types of elementary particles, with respect to spin:
scalars, spin one-half fermions and vector gauge bosons. The
quantum field theory for such types of particles has been
successfully constructed and has passed through severe
experimental examinations. Almost all phenomenological attempts to
extend the SM consist in introducing of a number of new particles
of the same type, like extra Higgs multiplets, fourth fermion
generation and various $Z'$ gauge bosons. Even SUSY phenomenology
does not go further simple doubling of known types of particles.

In this letter we would like to discuss again the possibility of
the presence in nature of new types of spin-1 chiral bosons, which
are complementary to the gauge ones. While the latter transform
according to the {\em real} representation (1/2,1/2) of the
Lorentz group, transformation properties of the former ones are
associated with the {\em inequivalent chiral} representations
(1,0) and (0,1). The chiral spin-1 bosons are described by
rank-two antisymmetric tensor fields $T_{mn}$, instead of vector
potential $V_m$, which describes the gauge spin-1 bosons.

On the mass shell both fields describe spin-1 particles and an
equivalence relation between them can be set in the massive case
\begin{equation}\label{equivalence}
  T_{mn}=\frac{1}{M}\left(\partial_m V_n-\partial_n V_m\right)-
  \frac{\epsilon_{mnab}}{2M}\left(\partial_a A_b-\partial_b A_a\right),
\end{equation}
where $M$ is the mass of spin-1 particles. The tensor field
describes simultaneously two three-component vector $V_m$ and
axial-vector $A_m$ fields.  They can be expressed through
six-component tensor field $T_{mn}$, using the on-shell operator
equation $\partial^2=-M^2$,
\begin{equation}\label{inverse}
  V_m=\frac{1}{M}\,\partial_n T_{mn},\hspace{0.5cm}
  A_m=\frac{1}{2M}\,\epsilon_{mnab}\,\partial_n T_{ab}.
\end{equation}
However, many years ago it was noted~\cite{Kemmer} already that
different Yukawa interactions for different types of spin-1 bosons
lead to inequivalent models. The first successful phenomenological
application of this idea has been realized for hadron meson states
with spin one~\cite{TNJL}.

It was pointed out that {\em three different} quantum numbers
$J^{PC}$ of existing spin-1 mesons, $1^{--}$, $1^{++}$ and
$1^{+-}$, cannot be assigned just to {\em two} vector
$\bar{q}\gamma_m q$ and axial-vector $\bar{q}\gamma_m \gamma^5 q$
quark states. So, the rank-two antisymmetric tensor current
$\bar{q}\sigma_{mn} q$ was introduced, which also describes vector
and axial-vector boson states, but with different transformation
properties with respect to Lorentz group and with different
quantum numbers $1^{--}$ and $1^{+-}$, respectively. This example
demonstrates that both the pure tensor states, $b_1$ bosons, and
mixed combinations of vector and tensor states, due to the same
quantum numbers, $\rho$ and $\rho'$ bosons, may exist.

These considerations, in general, support the idea of extension of
the SM with a new type of spin-1 chiral particles~\cite{MPL},
which has been proposed on pure theoretical grounds, in order to
describe the anomalies in the weak charged meson decays. Assuming
$SU_L(2)\times U_Y(1)$ symmetry of electroweak interactions and
that very heavy right-handed neutrinos $\nu_{Ra}$ are decoupled at
the weak scale, most general Yukawa interactions were proposed
\begin{eqnarray}\label{Y}
  {\cal L}_Y=&&\hspace{-0.3cm}\frac{1}{4}\left[
  t^q_a\left(\bar{Q}_a\sigma^{mn}d_{Ra}\right)+
  t^\ell_a\left(\bar{L}_a\sigma^{mn}e_{Ra}\right)
  \right]\left(
  \begin{array}{c} T^+_{mn} \\ T^0_{mn} \end{array}
  \right)
  \nonumber \\
  &&\hspace{-0.2cm}+\frac{u^q_a}{4}
  \left(\bar{Q}_a\sigma^{mn}u_{Ra}\right)
  \left(
  \begin{array}{c} U^0_{mn} \\ U^-_{mn} \end{array}
  \right)+{\rm h.c.},
\end{eqnarray}
where the two doublets $T_{mn}$ and $U_{mn}$ of the new tensor
chiral fields with opposite hypercharges $Y(T)=+1$, $Y(U)=-1$ were
introduced.

The fermion sector is the same as in the SM and consists of three
generations (noted by the index $a$) of the two component Weyl
spinors $\psi_a$: the left-handed lepton $L_a$ and the color quark
$Q_a$ doublets and the right-handed lepton $e_{Ra}$ and the color
quark $u_{Ra}$, $d_{Ra}$ singlets. Here and further on we omit
quark color indexes as far as the tensor fields are color blind.
The local gauge electroweak symmetry, maintained by the triplet
{\boldmath$W$}$\!_m$ and singlet $B_m$ bosons,  and $SU_C(3)$
color symmetry rest untouched. It was also argued~\cite{MPL} that
in order to avoid new chiral anomalies connected with the
antisymmetric chiral tensor fields, two Higgs doublets with
opposite hypercharges should be introduced instead of the one in
the SM.

Topologically the interactions (\ref{Y}) are the same as the
electroweak gauge interactions
\begin{equation}\label{G}
  {\cal L}_{\rm gauge}=
  g\left(\bar{\psi}_a\gamma^m \mbox{\boldmath$T$}\psi_a\right)
  \mbox{\boldmath$W$}\!_m+
  g'\left(\bar{\psi}_a\gamma^m \frac{Y}{2} \psi_a\right) B_m,
\end{equation}
where $g$, $g'$ are the gauge coupling constants, and
{\boldmath$T$}, $Y$ are the generators of $SU_L(2)$ and $U_Y(1)$,
correspondingly. The only principal difference between them
concerns fermion chiralities. While the gauge field interactions
preserve the helicities of incoming and outcoming fermions, the
tensor field couplings lead to a helicity flip. This feature
should be the main signature in their experimental search.

\section{Coupling constant relations}

In this section we will obtain relations among the gauge coupling
constants $g$, $g'$ and the tensor Yukawa couplings $t^\ell_a$,
$t^q_a$ and $u^q_a$. In order to do this we will apply hypothesis
about dynamical generation of kinetic terms for the bosons. This
idea comes from Nambu and Jona-Lasinio model~\cite{NJL} with
nonlinear four-fermion interactions. It has been developed in
\cite{Eguchi} for the more appropriate for quantization linear
form of interactions, introducing auxiliary boson fields (without
kinetic terms) with dimensionless Yukawa coupling
constants~\footnote{It is interesting to note that in the same
paper the author has made a conclusion about the impossibility to
introduce an antisymmetric tensor in a chiral-invariant manner. It
has been done in \cite{TNJL}.}.

Let us first apply this method for the gauge fields
{\boldmath$W$}$\!_m$ and $B_m$, assuming that they do not posses
initial kinetic terms. The kinetic terms are generated by
self-energy quantum corrections from fermionic loops. Subsequent
renormalization procedure leads to proper normalization of the
kinetic terms and sets relations among various Yukawa coupling
constants.

So, in the lowest one-loop approximation the self-energy quantum
corrections to the gauge fields {\boldmath$W$}$\!_m$ read
\begin{eqnarray}\label{W}
  &&\hspace{-0.5cm}\Pi_{W^{ij}_{mn}}(q)
  =ig^2(1+N_C)N_g
  ~{\rm Tr}\left(T^i T^j\right)
  \nonumber \\
  &&\times\int\!\!\frac{{\rm d}^4p}{(2\pi)^4} {\rm Tr}\left[\gamma_m
  \frac{1-\gamma^5}{2} (p\bk - q\bk)^{-1}
  \gamma_n \frac{1-\gamma^5}{2}
  p\bk^{-1}\right]
  \nonumber \\
  &&\hspace{1.2cm}=-4g^2 I_0\left(g_{mn}q^2-q_m q_n\right)\delta^{ij}
  +{\cal O}(q^4),
\end{eqnarray}
where $I_0=-i/(2\pi)^4{\rm Reg}\int{\rm d}^4p/p^4$ is the
regularized value of the logarithmically divergent integral.
Analogous calculations for the abelian gauge field $B_m$ give
\begin{equation}\label{B}
  \Pi_{B_{mn}}(q)=-\frac{20}{3}\,g'^2 I_0
  \left(g_{mn}q^2-q_m q_n\right)+{\cal O}(q^4).
\end{equation}

The comparison of the two previous equations leads to a relation
between the gauge coupling constants
\begin{equation}\label{gg'}
  \left(\frac{g'}{g}\right)^2=\frac{3}{5}
\end{equation}
and, hence, to a prediction of the electroweak mixing angle
$\sin^2\theta_W=3/8$. For a first time this result was obtained in
\cite{Terazava}. This value coincides with the prediction of the
GUT theories (which include the simple gauge group $SU(5)$) at the
unification scale~\cite{SU5}. Due to the different behavior of the
evolution of gauge coupling constants $g$ and $g'$ with energy,
the renormalization effects lead to a decreasing of
$\sin^2\theta_W$ at low energies~\cite{GQW}, bringing it closer to
the experimental value~\cite{PDG}
\begin{equation}\label{sin2}
  \sin^2\theta^{\rm eff}_W=0.23152(14).
\end{equation}

A lack of a gauge symmetry for the chiral tensor interactions
(\ref{Y}) does not allow to fix the Yukawa coupling constants
$t^\ell_a$, $t^q_a$ and $u^q_a$. Therefore, to go further we must
make model dependent assumptions. Let us assume lepton-quark
universality for the tensor interactions, which are also blind to
the generations $t^\ell_a=t^q_a=t$, $u^q_a=u$.

Before proceeding with calculations of self-energy quantum
corrections to the tensor fields, it is convenient to rewrite them
through the more common notations of the vector potential. This is
possible using extension of the relation (\ref{equivalence})
off-shell~\cite{TNJL}
\begin{equation}\label{offshell}
  T_{mn}=\left(\hat{\partial}_m V_n-\hat{\partial}_n V_m\right)-
  \frac{\epsilon_{mnab}}{2}
  \left(\hat{\partial}_a A_b-\hat{\partial}_b A_a\right),
\end{equation}
where $\hat{\partial}_m=\partial_m/\sqrt{-\partial^2}$. Then eq.
(\ref{Y}) takes the form
\begin{eqnarray}\label{YVA}
  {\cal L}_Y=&&\hspace{-0.3cm}t\left(
  \bar{Q}_a\sigma^{mn}d_{Ra}+\bar{L}_a\sigma^{mn}e_{Ra}
  \right)\left(
  \begin{array}{c} \hat{\partial}_m T^+_n \\
  \hat{\partial}_m T^0_n \end{array}
  \right)
  \nonumber \\
  &&\hspace{-0.5cm}+u\left(\bar{Q}_a\sigma^{mn}u_{Ra}\right)
  \left(
  \begin{array}{c} \hat{\partial}_m U^0_n \\
  \hat{\partial}_m U^-_n \end{array}
  \right)+{\rm h.c.},
\end{eqnarray}

Now it is easy to calculate the self-energy quantum corrections to
doublets $T_m$
\begin{equation}\label{T}
  \Pi_{T_{mn}}(q)=-4t^2 I_0\left(g_{mn}q^2-q_m q_n\right)
  +{\cal O}(q^4)
\end{equation}
and $U_m$
\begin{equation}\label{U}
  \Pi_{U_{mn}}(q)=-3u^2 I_0\left(g_{mn}q^2-q_m q_n\right)
  +{\cal O}(q^4).
\end{equation}
The comparison of (\ref{W}), (\ref{T}) and (\ref{U}) leads to the
relation
\begin{equation}\label{relation}
  g^2=t^2=\frac{3}{4}\,u^2.
\end{equation}
An asymmetry in the coupling constants for the doublets of the
tensor fields arises due to their obviously non-symmetric
interactions to the leptons. At high energy scale, when the
right-handed neutrinos become active, the symmetry between both
interactions should be restored.

It may be expected that the evolution of the coupling constants
towards low-energy scale should not drastically change the
relation (\ref{relation}), since both the gauge coupling constant
$g$ and the Yukawa coupling constants $t$ and $u$ have
asymptotically free behavior~\cite{Avdeev}. We will use this
relation below for a quantitative determination of the masses of
the chiral particles.

\section{Masses and mixings}

The Fermi theory of the weak interaction~\cite{Fermi} gives
excellent example of effective field theory which perfectly
describes low-energy physics with only one dimensional coupling
constant $G_F=1.16637(1)\times 10^{-5}$~GeV$^{-2}$. It arises from
the exchange of the charged massive intermediate bosons $W^{\pm}$
at small momentum transfer, which at the tree level is expressed
as
\begin{equation}\label{GF}
  \frac{G_F}{\sqrt{2}}=\frac{g^2}{8M^2_W},
\end{equation}
where $M_W$ is the boson mass. In the SM the gauge coupling
constant $g^2=4\pi\alpha/\sin^2\theta_W$ is related to the
electromagnetic fine-structure constant $\alpha$ and
$\sin^2\theta_W$. It allows to estimate $M_W\simeq 77.5$~GeV at
the tree level with relatively good precision.

Therefore, in order to get a constraint on the mass of the new
intermediate boson one can analyze the precision low-energy
experiments, looking for an eventual admixture of new forces in
the ordinary weak processes. Particularly, the chiral bosons can
enhance the rates of the chirally suppressed decay modes.

One may think that $\pi\to e\nu$ decay is the ideal process to put
constraints on the presence of new currents chirally different
from the SM ones. However, for the tensor current the hadron
matrix element $\langle 0\vert \bar{q}\sigma_{mn}
q\vert\pi\rangle$ is zero by kinematic reasons, and this decay can
restrict only the presence of pseudoscalar
currents~\cite{Shankar}, but not tensor ones. Nevertheless, the
quark tensor currents can contribute to the rare three-body pion
decay $\pi\to e\nu\gamma$, where huge experimental anomalies have
been detected~\cite{Bolotov}.

In order to explain these anomalies, the following effective
tensor interactions have been derived~\cite{MPL}
\begin{eqnarray} \label{eff}
{\cal L}^{\rm eff}_T=\hspace{-0.2cm}&-&\hspace{-0.2cm}\sqrt{2}f_T
G_F\,\bar{u}\sigma_{ml}d_L\, \frac{4q^l q_n}{q^2}\,
\bar{e}\sigma^{mn}\nu_L
\nonumber\\
&-&\hspace{-0.2cm}\sqrt{2}f'_T G_F\,\bar{u}\sigma_{ml}d_R\,
\frac{4q^l q_n}{q^2}~ \bar{e}\sigma^{mn}\nu_L+{\rm h.c.},
\end{eqnarray}
where $q$ is the momentum transfer between quark and lepton pairs
and $f_T$, $f'_T$ are dimensionless coupling constants, which
determine the strength of the new tensor interactions relative to
the ordinary weak interactions.

These interactions arise from the exchange of the charged chiral
$T^\pm$ and $U^\pm$ bosons at low energies. The first term in
(\ref{eff}) is generated due to the non-diagonal elements of mass
mixing matrix~\cite{MPL}
\begin{equation}\label{MTU}
  {\cal M}^2=\left(\begin{array}{cc} T^+_m & U^+_m
  \end{array}\right) \left(
  \begin{array}{cc}
    M^2 & -\mu^2 \\
    -\mu^2 & m^2
  \end{array}\right)
  \left(\begin{array}{c}
    T^-_m \\ U^-_m
  \end{array}\right).
\end{equation}
It can be rewritten identically in a more common form without
momentum dependence
\begin{equation}\label{identity}
  \bar{u}\sigma_{ml}d_L~
  \frac{4q^l q_n}{q^2}~ \bar{e}\sigma^{mn}\nu_L
  \equiv
  \bar{u}\sigma_{mn}d_L~\bar{e}\sigma^{mn}\nu_L.
\end{equation}

This local tensor interaction has been considered early in $\beta$
decay phenomenology on the same footing as other local
four-fermion interactions. However, to my knowledge, the second
non-local term in (\ref{eff}) has never been used before 1993.
Owing to the identity $\bar{u}\sigma_{mn}
d_R~\bar{e}\sigma^{mn}\nu_L\equiv 0$ it is impossible to construct
{\em local} tensor interaction with an opposite to
(\ref{identity}) chirality of quark current.

It is worth noting, that the two terms of the tensor interactions
(\ref{eff}) arise as a result of the chiral boson exchanges and
should be considered simultaneously. The requirement of positivity
of the determinant $\Delta=M^2 m^2-\mu^4>0$ of the square mass
matrix (\ref{MTU}) leads to the positive effective coupling
constants
\begin{equation}\label{fT}
  f_T=\frac{g^2 m^2}{4\sqrt{2}G_F\Delta}>0,\hspace{1cm}
  f'_T=\frac{g^2 \mu^2}{2\sqrt{6}G_F\Delta}>0
\end{equation}
and the pseudotensor quark currents $\bar{u}\sigma^{mn}\gamma^5 d$
can completely cancel out in (\ref{eff}) in the case $f_T=f'_T$.
This cancellation helps to avoid another constraint on the tensor
interaction (\ref{identity}) following from the pion decay due to
generation of the pseudoscalar quark current $\bar{u}\gamma^5 d$
through the electromagnetic radiative corrections~\cite{Voloshin}.

In the case of equality of the effective tensor coupling
constants, the diagonalization of the mass matrix (\ref{MTU})
gives two mass states
\begin{equation}\label{MT}
  M^2_{H/L}=\frac{M^2+m^2\pm\sqrt{\left(M^2-m^2\right)^2+3m^4}}{2}.
\end{equation}
The maximum value of the lightest mass state is reached at
$M^2=2m^2$ and defines the physical mass $M^2_L=m^2/2$, which
corresponds to an energetically favored exchange by this particle.
Then, the heavy state has a mass $M^2_H=5M^2_L$.

The centi-weak tensor interaction $f_T\approx 10^{-2}$ can explain
the deficit of events in the radiative pion decay and, assuming
$e-\tau$ universality, the excess in $\tau\to 2\pi\nu$ decay in
comparison with CVC prediction~\cite{tau}. This value of the
effective tensor coupling constant allows to predict the masses of
the charged chiral bosons
\begin{equation}\label{MHL}
  M_H=\sqrt{\frac{2}{f_T}}\,M_W\approx 1137~{\rm GeV},
  \hspace{0.3cm} M_L\approx 509~{\rm GeV},
\end{equation}
which will be used in the next section for quantitative estimations of
their production cross-sections.

These physical massive states are represented by two orthogonal
combinations ${\cal U}^\pm_m=(\sqrt{3}\,U^\pm_m+T^\pm_m)/2$ and
${\cal T}^\pm_m=(\sqrt{3}\,T^\pm_m-U^\pm_m)/2$, which correspond
to light and heavy massive particles, correspondingly. Their
Yukawa interactions take the form
\begin{eqnarray}\label{YC}
    {\cal L}^C_Y=&&\hspace{-0.3cm}
    \frac{g}{2}\left(
    \bar{u}_a\sigma^{mn}d_{Ra}+
    \bar{\nu}_a\sigma^{mn}e_{Ra}\right)
    \left(\hat{\partial}_m{\cal U}^+_n+
    \sqrt{3}\,\hat{\partial}_m{\cal T}^+_n\right)
    \nonumber \\
    &&\hspace{-0.4cm}+\,g\left(\bar{u}_a\sigma^{mn}d_{La}\right)
    \left(\hat{\partial}_m{\cal U}^+_n-\frac{1}{\sqrt{3}}\,
    \hat{\partial}_m{\cal T}^+_n\right)+{\rm h.c.}
\end{eqnarray}

Besides the mixing (\ref{MTU}) between the charged chiral
particles themselves, they can mix with the $W^\pm$ boson leading
to anomalous moment for the charged weak transitions~\cite{kW}.
The first experimental constraints on this parameter have been
obtained in \cite{DELPHI} from $\tau$-lepton decays. The mixings
between the gauge and the chiral bosons are in a complete analogy
with $\rho-\rho'$ mixing in hadron physics~\cite{TNJL}. However,
in contrast to the latter, where the mixing is maximal,
experimental data point out that the weak mixing is rather small.

Let us consider the neutral spin-1 sector of the model. Here
besides electroweak bosons $\gamma$ and $Z$ we have introduced
{\em four} degrees of freedom
$T^R_m=(T^0_m+\bar{T}^0_m)/\sqrt{2}$,
$T^I_m=(T^0_m-\bar{T}^0_m)/\sqrt{2}i$,
$U^R_m=(U^0_m+\bar{U}^0_m)/\sqrt{2}$,
$U^I_m=(U^0_m-\bar{U}^0_m)/\sqrt{2}i$ for the neutral chiral
bosons. Due to their quantum numbers $T^I_m$ and $U^I_m$ states do
not mix and decouple from the others in the case of $CP$ invariant
interactions. Although $T^R_m$ and $U^R_m$ states do not mix
between each other, because they couple to the different types
$up$ and $down$ fermions, they can mix with the photon and $Z$
boson leading to additional contributions to anomalous magnetic
and neutral weak moments for the fermions. We will assume, in
agreement with the experimental data, that these mixings are very
small. Therefore, one can consider these states as physical with
the following Yukawa interactions
\begin{eqnarray}\label{YN}
  {\cal L}_Y^N=&&\hspace{-0.3cm}
    \frac{g}{\sqrt{2}}\left(
    \bar{d}_a\sigma^{mn}d_a+
    \bar{e}_a\sigma^{mn}e_a\right)
    \hat{\partial}_m T^R_n
    \nonumber \\
    &&\hspace{-0.7cm}+\,i\frac{g}{\sqrt{2}}\left(
    \bar{d}_a\sigma^{mn}\gamma^5d_a+
    \bar{e}_a\sigma^{mn}\gamma^5e_a\right)
    \hat{\partial}_m T^I_n
    \nonumber \\
    &&\hspace{-0.7cm}+\sqrt{\frac{2}{3}}\,g\left(
    \bar{u}_a\sigma^{mn}u_a\right)
    \hat{\partial}_m U^R_n
    \nonumber \\
    &&\hspace{-0.7cm}+\,i\sqrt{\frac{2}{3}}\,g\left(
    \bar{u}_a\sigma^{mn}\gamma^5u_a\right)
    \hat{\partial}_m U^I_n
\end{eqnarray}

In order to estimate the masses of the neutral chiral bosons we
can use the mass matrix (\ref{MTU}). Since the neutral chiral
bosons belong to the same multiplets as their charged partners and
they do not mix, their masses can be estimated as
\begin{eqnarray}\label{M0}
  M_U&=&m=\sqrt{2}M_L\approx 719~{\rm GeV},
  \nonumber\\
  M_T&=&M=2M_L\approx 1017~{\rm GeV}.
\end{eqnarray}
Degeneracies in the masses of $U^I_m$, $U^R_m$ and $T^I_m$,
$T^R_m$ states are removed as a result of mixings of $U^R_m$ and
$T^R_m$ states with the photon and $Z$ boson, leading to the
inequalities $M_{U^R}>M_{U^I}=M_U$ and $M_{T^R}>M_{T^I}=M_T$.

It is a hard task to detect neutral tensor interactions on the
background electromagnetic and neutral weak processes in low
energy experiments. Up to now there are no positive indications,
besides, may be, the deviation in the muon anomalous magnetic
moment, which can be assigned to mixings between the photon and
the neutral chiral bosons.
Nevertheless, as we will show in the next section the chiral
bosons have a unique signature for their detection at hadron
colliders due to their anomalous couplings with fermions.

\section{Results and Conclusions}

The main interest represents the production cross-sections of the
lightest chiral bosons. In order to estimate them, we will
calculate basic partial decay widths of the chiral bosons into
leptons and quarks.

Using the Yukawa interactions (\ref{YC}) we can evaluate the
lepton
\begin{equation}\label{Gl}
    \Gamma_\ell\equiv\Gamma({\cal U}\to\ell\nu)=
    \frac{g^2 M_L}{192\pi}=\frac{\Gamma(W\to\ell\nu)}
    {\sqrt{40f_T}} \approx 360~{\rm MeV}
\end{equation}
and the quark
\begin{equation}\label{Gq}
    \Gamma_q\equiv\Gamma({\cal U}\to\bar{u}d)=15\Gamma_\ell\approx 5.4~{\rm GeV}
\end{equation}
widths of the charged chiral bosons ${\cal U}^\pm$. Assuming the
presence of only three fermion generations with masses lighter
than $M_L$ we can estimate the total width as
$\Gamma=48\Gamma_\ell\approx 17.2$~GeV.

Then it is possible to evaluate the production cross-sections at
the Tevatron~\footnote{Since the ${\cal U}$ cross section is
mainly determined by a valence quark from the proton and an
anti-quark from the anti-proton, we neglect here the contribution
of the sea quarks.}
\begin{eqnarray}\label{sTev}
    \sigma^{\rm Tev}_{{\cal U}^\pm}&=&
    \frac{4\pi^2\Gamma_q\tau}{3M^3_L}\!
    \int^1_\tau\hspace{-0.2cm} u(x,M_L)d\left(\frac{\tau}{x},M_L\right)
    \!\frac{{\rm d}x}{x}
    \nonumber\\
    &\approx& \left\{\begin{array}{ll}
      8.4~{\rm pb} & {\rm RUN~I} \\
      11.7~{\rm pb} & {\rm RUN~II} \
    \end{array}\right.
\end{eqnarray}
and the LHC
\begin{eqnarray}\label{sLHC}
    &&\hspace{-0.7cm}
    \sigma^{\rm LHC}_{{\cal U}^+}\!\!=
    \frac{8\pi^2\Gamma_q\tau}{3M^3_L}\!\!
    \int^1_\tau\hspace{-0.3cm} u(x,M_L)\bar{d}\left(\frac{\tau}{x},M_L\right)
    \!\frac{{\rm d}x}{x}
    \approx 0.36~{\rm nb},
    \nonumber \\
    &&\hspace{-0.7cm}
    \sigma^{\rm LHC}_{{\cal U}^-}\!\!=
    \frac{8\pi^2\Gamma_q\tau}{3M^3_L}\!\!
    \int^1_\tau\hspace{-0.3cm} \bar{u}(x,M_L)d\left(\frac{\tau}{x},M_L\right)
    \!\frac{{\rm d}x}{x}
    \approx 0.19~{\rm nb}
\end{eqnarray}
using, for example, CTEQ6M parton distribution
functions~\cite{CTEQ6} and the appropriated parameter
$\tau=M^2_L/s$, where $s$ is the square of the center-of-mass
energy.

Such a big production cross-sections would immediately contradict
to the present Tevatron data, unless the chiral bosons have
unusual properties. It is noteworthy, that the lepton width
(\ref{Gl}) is suppressed in favor of the quark one (\ref{Gq}).
Hence, the lightest charged chiral particle shows leptophobic
property that leads to a reduced lepton production
\begin{eqnarray}\label{sTev_l}
    \hspace{-0.2cm}&&\sigma^{\rm Tev}_{\ell^\pm}=
    \sigma^{\rm Tev}_{{\cal U}^\pm}
    \times B({\cal U}\to\ell\nu)
    \approx\left\{
    \begin{array}{ll}
      0.18~{\rm pb} & {\rm RUN~I} \\
      0.24~{\rm pb} & {\rm RUN~II} \
    \end{array}\right.,
    \nonumber \\
    \hspace{-0.3cm}&&\sigma^{\rm LHC}_{\ell^+}\approx 7.5~{\rm pb},
    \hspace{0.7cm}
    \sigma^{\rm LHC}_{\ell^-}\approx 4~{\rm pb},
\end{eqnarray}
which is the main channel of charged boson detection at hadron
colliders.

However, the cross sections (\ref{sTev_l}) are still big to remain
undetected at the Tevatron. This fact can be explain by another
unusual and unexpected feature of the chiral bosons connected to
their anomalous interactions (\ref{YVA}) with fermions. Let us
compare the normalized angular distributions of the lepton from
the decays of $W^-$
\begin{equation}\label{NW}
    \frac{{\rm d}N_W}{{\rm d}\Omega}=\left\{
    \begin{array}{ll}
    \frac{3}{16\pi}(1\mp\cos\theta)^2, & \lambda=\pm1,\\
    &\\
    \frac{3}{8\pi}\sin^2\theta, & \lambda=0,
    \end{array}\right.
\end{equation}
and the ${\cal U}^\pm$
\begin{equation}\label{NU}
    \frac{{\rm d}N_{\cal U}}{{\rm d}\Omega}=\left\{
    \begin{array}{ll}
    \frac{3}{8\pi}\sin^2\theta, & \lambda=\pm1,\\
    &\\
    \frac{3}{4\pi}\cos^2\theta, & \lambda=0,
    \end{array}\right.
\end{equation}
where $\lambda$ is the boson helicity.

For example, the left-handed quark $d$ (from the proton)
interacting with the right-handed anti-quark $\bar{u}$ (from the
anti-proton) can produce the $W^-$ with spin projection on the
proton beam direction $-1$. Hence, the decay leptons are
distributed as $(1+\cos\theta)^2$. While chiral particle
production arises from the interaction of a quark and an
anti-quark with the same helicities, that leads to zero helicity
of the produced chiral boson and $\cos^2\theta$ lepton
distribution~\cite{two}.

Indeed, the cross section for $p+\bar{p}\to {\cal U}+X\to\ell+X$
process
\begin{equation}\label{S}
    {\rm d}\sigma=\frac{1}{3}
    \int\hspace{-0.1cm}{\rm d}x_1{\rm d}x_2\,
    u(x_1,M_L)d(x_2,M_L)\,
    {\rm d}\hat{\sigma}(\hat{s},\hat{t})
\end{equation}
is expressed through the relevant differential cross section of
the parton subprocess $d+\bar{u}\to{\cal U}^-\to\ell+\bar{\nu}$
\begin{equation}\label{s}
    E_\ell\frac{{\rm d}^3\hat{\sigma}(\hat{s},\hat{t})}
    {{\rm d}^3 p_\ell}=
    \frac{5g^4}{(32\pi)^2}\,
    \frac{(\hat{s}+2\hat{t}\,)^2\delta(\hat{s}+\hat{t}+\hat{u})}
    {\hat{s}|\hat{s}-M_L^2+iM_L\Gamma|^2},
\end{equation}
where $\hat{s}=(p_d+p_{\bar{u}})^2$,
$\hat{t}=(p_{\bar{u}}-p_\ell)^2$ and $\hat{u}=(p_d-p_\ell)^2$ are
the Mandelstam variables. In the center-of-mass parton system the
differential cross section shows the following distribution
\begin{equation}\label{cos2}
    \frac{{\rm d}\hat{\sigma}}{{\rm d}\Omega}\propto
    (\hat{s}+2\hat{t}\,)^2\propto
    \cos^2\hat{\theta}=1-\frac{4\hat{p}^2_T}{\hat{s}}.
\end{equation}
Here $\hat{\theta}$ is the angle between the lepton and the parton
direction and $\hat{p}^2_T$ is the square of the transverse lepton
momentum.

Since the latter is invariant under longitudinal boosts along the
beam direction, the distribution (\ref{cos2}) versus $\hat{p}_T$
holds also in the lab frame $p_T=\hat{p}_T$. Changing variables in
the differential cross section from $\cos\hat{\theta}$ to $p^2_T$
\begin{equation}\label{Jacobian}
    \frac{{\rm d}\cos\hat{\theta}}{{\rm d}p^2_T}=
    -\frac{2}{\hat{s}}\left(\sqrt{1-\frac{4p^2_T}{\hat{s}}}\,\right)^{-1}
\end{equation}
leads to a kinematical singularity at the endpoint
$p^2_T=\hat{s}/4$, which gives the prominent Jacobian peak in the
$W$ decay distribution.

\begin{figure}[th]
\epsfig{file=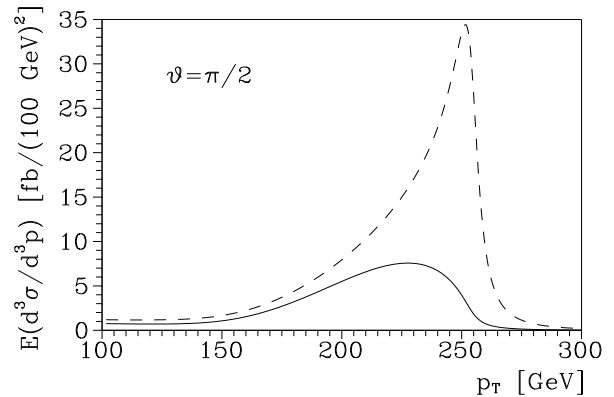,height=5.3cm,width=8cm}
\caption{\label{fig:1} Lepton momentum distributions transverse to
the beam axis at Tevatron. Solid curve shows the ${\cal U}$-decay
distribution, while dashed line correspond to the gauge-like boson
distribution, where $\cos^2\hat{\theta}$ in (\ref{cos2}) is
replaced with $\frac{1}{4}(1+\cos\hat{\theta})^2$.}
\end{figure}

In contrast to this, the pole in the decay distribution of the
chiral bosons is cancelled out and, moreover, the distribution
reaches zero at the endpoint $p^2_T=\hat{s}/4$. The chiral boson
decay distribution has a broad smooth bump with the maximum below
the kinematical endpoint, instead of sharp Jacobian peak
(Fig.~\ref{fig:1}). In the case when the chiral boson is produced
with no transverse momentum, the transverse mass of the lepton
pair is related to $p_T$ as $M_T(\ell\nu)=2p_T$ and the Jacobian
peak is absent in $M_T$ distribution as well.

Therefore, the decay distribution of the chiral bosons differs
drastically from the distribution of the gauge bosons. Even
relatively small decay width of the chiral bosons leads to a wide
distribution, that obscures their identification as resonances at
hadron colliders. At the present time one can speak only about
eventual excess of events in the region $350$ GeV $<M_T<500$ GeV,
where the background from the tail of the $W$ decays is
considerably small. Indeed, such excess, although not
statistically significant, has been pointed out recently by the
CDF Collaboraion~\cite{Wp} in the same region.

The form of the decay distribution for the chiral bosons resembles
the bump anomalies in the inclusive jet $E_T$ distribution,
reported by the CDF Collaboration~\cite{CDF}. It has been
suggested that these anomalies could be described by a new neutral
leptophobic gauge boson with huge decay width~\cite{leptophobia}.
Although this problem has been solved in the framework of the SM
by changing the gluon distribution functions~\cite{gluon}, it
could be reconsidered in the light of the new form of the decay
distribution as a real physical signal from decays of different
chiral bosons, both charged and neutral.

Analysing the bumps in the jet transverse energy distribution in
Fig.~1 of ref. \cite{CDF}, we can find the endpoint of the first
bump at 250 GeV and guess about the second bump endpoint from the
minimum around 350 GeV. If we assign the first bump to the hadron
decay products of the lightest charged bosons, which exactly
corresponds to the estimated mass from eq. (\ref{MHL}), the second
endpoint hints to a mass for the lightest neutral boson around
700~GeV, which is also in a qualitative agreement with our
estimations (\ref{M0}). However, taking into account the large
systematic uncertainties in jet production, these conclusions may
be premature.

Nevertheless, the lightest neutral chiral bosons $U^R_m$ and
$U^I_m$ can fairly play the role of the hadrophilic boson,
suggested in \cite{leptophobia} in order to explain Tevatron and
LEP anomalies, with nearly the same mass. Indeed, these bosons do
not decay into charged leptons and cannot be seen in lepton
channel. In the best case they have invisible decay widths in
neutrinos, if right-handed neutrinos are not too heavy. However,
in contrast to \cite{leptophobia} these bosons are non-gauge
bosons, but chiral ones. There exist also additional neutral
chiral bosons $T^R_m$ and $T^I_m$, which couple only to the {\em
down} fermions with approximately the same Yukawa coupling
constants as $U^R_m$ and $U^I_m$ bosons. They influence the
low-energy hadron physics two times weaker than the $U^R_m$ and
$U^I_m$ bosons due to their higher masses.

The production of the heaviest neutral chiral bosons $T^R_m$,
$T^I_m$ and the detection of their leptonic decays at the hadron
colliders will be crucial test of the discussed model.
Unfortunately, due to their big masses they cannot be produced at
the Tevatron at present. Let us calculate their lepton and quark
widths in order to estimate their production cross-section at the
LHC.

Using the Yukawa interactions (\ref{YN}) we obtain the following
decay widths
\begin{equation}\label{Tl}
  \Gamma^0_\ell\equiv\Gamma(T^0\to\ell^+\ell^-)=
    \frac{g^2 M_T}{48\pi}\approx 2.9~{\rm GeV}
\end{equation}
for the lepton and
\begin{equation}\label{Tq}
    \Gamma^0_q\equiv\Gamma(T^0\to\bar{d}d)=
    3\Gamma^0_\ell\approx 8.6~{\rm GeV}
\end{equation}
for the quark channels. This gives $\Gamma^0_T=12\Gamma^0_\ell
\approx 34$~GeV for the total fermion decay width. Since the
lepton branching ratio $B(T^0\to\ell^+\ell^-)=1/12$ is not
negligible one expects essential signal in the dilepton channels
with a cross section
\begin{eqnarray}\label{s2LHC}
    \sigma^{\rm LHC}_{\ell^+\ell^-}&=&\sigma^{\rm LHC}_{T^0}\times
    B(T^0\to\ell^+\ell^-)
    \nonumber \\
    &&\hspace{-1.5cm}=\frac{8\pi^2\Gamma^0_q\Gamma^0_\ell\tau}{3M^3_T\Gamma^0_T}\!\!
    \int^1_\tau\hspace{-0.3cm} d(x,M_T)\bar{d}\left(\frac{\tau}{x},M_T\right)
    \!\frac{{\rm d}x}{x}
    \approx 1~{\rm pb}.
\end{eqnarray}

The most conclusive experimental evidence for the presence of
anomalous interactions and new type of spin-1 non-gauge bosons
will be the detection of wide bumps in leptonic distribution at
high $p_T$ at Tevatron and LHC and the dilepton production with
high invariant mass around 1 TeV. Considering $e^\pm$ and
$\mu^\pm$ channels for the decays of the lightest charged chiral
bosons, one may expect around 1 event per day at Tevatron (average
luminosity $10^{31}$ cm$^{-2}$s$^{-1}$) and around 1 event every 5
seconds at LHC (designed peak $L=10^{34}$ cm$^{-2}$s$^{-1}$).
Taking into account also that in $d\bar{d}$-annihilation two
different species of the neutral chiral bosons $T^R_m$ and $T^I_m$
are produced with approximately the same masses, one may expect
one electron or muon pair with invariant mass around 1~TeV every
25 second at LHC peak luminosity.

In conclusion, we would like to note that our numerical results
essentially depend on the magnitude of the low-energy effective
tensor coupling constant $f_T$. According to our
estimations~\cite{discovery} its value can be up to 30\% larger,
which effectively decreases all the calculated masses of the
chiral bosons by 15\%. However, the mass ratios of the chiral
bosons
\begin{equation}\label{Mratio}
    M_L:M_U:M_T:M_H=1:\sqrt{2}:2:\sqrt{5}
\end{equation}
are fixed by the model. Really, the masses of the charged ${\cal
U}^{\pm}_m$, ${\cal T}^{\pm}_m$ and neutral $U^R_m$, $T^R_m$
chiral bosons can get an additional positive contributions due to
the small mixings between the gauge and the chiral bosons, which
have been neglected here.

Taking into account the unusual lepton distribution (\ref{cos2}),
the main attention should be paid to events with large
pseudorapidities, where the dominant lepton production occurs as a
result of decays of the new chiral particles. To support these
qualitative statements and to be ready to meet eventual new
physics, detailed quantative calculations should be fulfilled on
the basis of the interactions (\ref{YC}) and (\ref{YN}), taking
into account also all possible known effects and specifics of
concrete detector.

\section*{Acknowledgements}

I am grateful to D. Kirilova for the overall help and appreciate
the stimulating ICTP environment.

\pagebreak[3]

\end{document}